# Mathematical modeling of hysteresis loops in ferroelectric materials

Zhi MA*, Liying XI, Xuming WANG, Yanan MA, Fu ZHENG, Hua GAO, Hongfei LIU, Huanming CHEN

*School of Physics & Electronic-Electrical Engineering, Ningxia University, Yinchuan 750021, PR China*

*Corresponding author: mazhicn@126.com (Zhi MA)

**Abstract.** In published papers, the Gibbs free energy of ferroelectric materials has usually been quantified by the retention of 6th or 8th order polarization terms. In this paper, a newly analytical model of Gibbs free energy, thereout, a new model of polarization-electric field hysteresis loops in ferroelectric materials has been derived mathematically. As a model validation, four patterns of polarization-electric field hysteresis loops of ferroelectric materials have been depicted by using the model. The calculated results indicated that the self-similar model can characterize the various patterns of hysteresis loops in ferroelectric materials through adjusting the external excitation or the synthetically parameter (e.g., electric, temperature, and stress, etc.) employed in the model.

## 1. Introduction

Due to the intrinsic properties of the ferroelectrics, such as coercive field, spontaneous polarization, and remnant polarizations can be extracted from the polarization-electric field hysteresis loops directly, the polarization-electric field hysteresis phenomena of ferroelectric materials have been investigated extensively since the ferroelectric phenomena being discovered[1-5]. Experimentally, the hysteresis loops can be measured while a periodical electric field or stress being applied onto the ferroelectrics externally[6-7]. Furthermore, many factors, such as temperature, grain boundaries or phase boundaries, doping, and anisotropy etc., which can affect the hysteretic behaviors of the ferroelectrics, have also been studied extensively[8-12]. Theoretically, some effects of microstructure and applied strain or stress taking on the hysteretic behaviors of the ferroelectrics have been predicted for many years similarly, including investigation of the grain size effect on the hysteresis properties of $BaTiO_3$ polycrystals, the effect of dislocation walls on the polarization switching of a ferroelectric single crystal, the effect of strain on the domain switching of ferroelectric polycrystals, and the strain effect in the problem of critical thickness for ferroelectric memory, etc..[13-18]. Many experimental and theoretical results indicated that the hysteretic behaviors of the ferroelectrics are not only dependent on the microstructure of materials (such as grain size, thickness, aging, and grain/phase boundaries) but also dependent on the measurement conditions strongly (such as frequency, temperature and stress etc.)[19-21]. Therefore, the reported hysteresis loops measured or predicted under different conditions inevitably presented with diversified patterns. And The diversified patterns obtained experimentally are still beyond on the predictions simulated by using the phenomenological theory of time-dependent Landau Ginzburg Devonshire(TDLGD)[17,18,22].

More recently, Li Jin, Fei Li and Shujun Zhang summarized the experimental hysteresis loops and categorized them into four groups based on their morphologic features. They also discussed the impact factors on the hysteresis loops[20]. In order to meet with and to predict the diversified patterns of hysteresis loops obtained from experiments theoretically, a new model of hysteresis loops in ferroelectric materials has been derived mathematically in this article. The effects of the external excitation, and the synthetically parameter (e.g., electric, temperature, and stress, etc.) employed in the model have been analyzed numerically, and it will be suggested that the model is feasible to describe the diversified patterns of hysteresis loops. To the authors' knowledge, the unified model proposed for predicting the diversified patterns of hysteresis loops in ferroelectric materials is first reported.

## 2. Mathematical Model of hysteresis in ferroelectric materials

According to the theory of fractal, hysteresis loops or the approximately reverse "*S*" curves obtained from experiments possess the characteristics of fractal. Therefore, we assume that the mathematical model of hysteresis loop can enable us to express its self-similarity. Naturally, based on the notions of self-similar system, let us start from a homogeneous function in general sense:

$$f(x) = \ln[\cosh(x)] = \ln\left(\frac{e^x + e^{-x}}{2}\right) \tag{1}$$

where $e$ is the base of the natural logarithm. In order to meet with the saturation for both asymptotic values of the externally applied field ($\vec{E}_{app} \to \pm\infty$), as well as the behavior of polarization, a differential from the equation (1) is as follow:

$$\frac{df(x)}{dx} = \tanh(x) = \frac{e^x - e^{-x}}{e^x + e^{-x}} = g(x) \tag{2}$$

In order to control and assure the scaling invariant, the function $g(x)$ has to satisfy the following conditions:

$$g(\lambda^\alpha x) = \lambda^\beta g(x) \tag{3}$$

where $\alpha$, $\beta$, and $\lambda$ are the scaling exponents.

We applied the equations mentioned above into the ferroelectric system. The equation (4) describing the polarization-electric field hysteresis loops of the ferroelectric system can be constructed, in which the vector $\vec{P}$ is corresponding to the spontaneous polarization, $T$ is temperature, $\sigma$ stands for the stress, and $k$ is polarization corresponding to saturation:

$$F_{\vec{P}}(\vec{P},T,\sigma) = kF(\vec{P},T,\sigma) = k\ln\{\cosh[(aT+b\sigma+c)\vec{P}+d]+h\} \tag{4}$$

where, we arbitrarily introduced a linearly action between the spontaneous polarization and the external excitation. For instance, the term $aT$ is the influence of temperature, $b\sigma$ is the linearly contribution of stress, etc. This was inspired by reference [23] which suggested that there is an equivalent effect of the electrical field, the mechanical stress, and the temperature taking on hysteresis loop. The parameters $a$, $b$, $c$, $d$ and $h$ are the related parameters during experimental measurement. In order to understand the role of the related parameters during experimental measurement, such as $k$, $d$ and $h$, clearly, we have the derivative of equation (4) with respect to the $\bar{P}$ and it can be expressed as equation (5):

$$G(\bar{P},T,\sigma) = k \cdot \Re \cdot \frac{\sinh(\Re \cdot \bar{P} + d)}{h + \cosh(\Re \cdot \bar{P} + d)} \tag{5}$$

Here, we set $\Re = aT + b\sigma + c$. The figures of equation (5) plotted under different combination of $k$, $d$ and $h$ reflected the basic features of the envelope of the hysteresis loop. It renders that the equation (5) can describe the hysteresis loops of ferroelectric materials. The various patterns of hysteresis loops in ferroelectric materials can be characterized through adjusting the scaling parameters, the external excitation, or the synthetically parameter (e.g., electric, temperature, and stress, etc.) employed in the model (Fig.1). Definitely, we correspond the saturation of polarization, the coercive field, and the slope of the hysteresis loop to the parameters of $k$, $d$ and $h$ respectively under a given temperature $T$ and a given stress $\sigma$. The model of hysteresis loop based on the equation (5) is invariant with respect to scaling and gauge transformation[24], which enables us to express its self-similarity by the homogeneous function in general sense and reproduce all polarization processes inside a major loop.

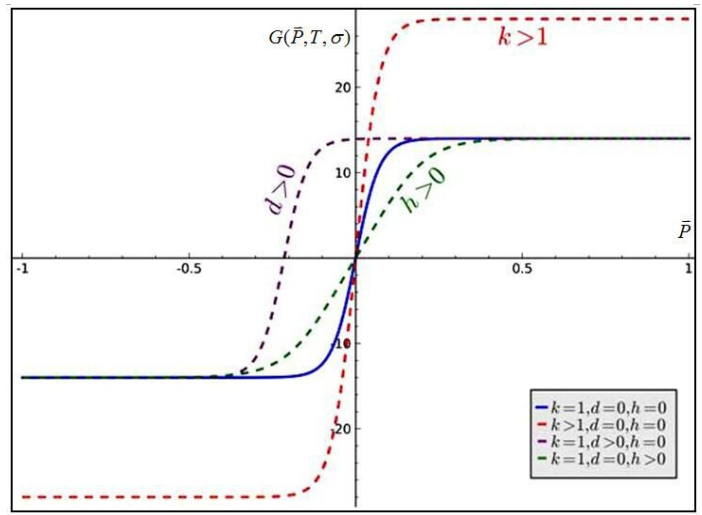

Fig.1 Dependence of $G(\bar{P},T,\sigma)$ on $\bar{P}$ under different combination of $k$, $d$ and $h$

## 3. Validation of the model

For simplify, we set $k=1$, $d=h=0$ and $\Re = aT+b\sigma+c$. The equation (4) and equation (5) can be rewritten as

$$F_{\bar{P}}(\bar{P},T,\sigma) = \ln[\cosh(\Re \cdot \bar{P})] \tag{4'}$$

$$G(\bar{P},T,\sigma) = \Re \cdot \tanh(\Re \cdot \bar{P}) \tag{5'}$$

We expanded equation (4)′ as a series:

$$F_{\bar{P}}(\bar{P}) = A\bar{P}^2 + B\bar{P}^4 + C\bar{P}^6 + D\bar{P}^8 + O(\bar{P}^{10}) \tag{6}$$

where $A$, $B$, $C$, and $D$ are the coefficients of the expansion. Apparently, the expanded series, equation (6), has the same form and is equivalent to the Landau free energy in ferroelectric materials[14,17,22]. Therefore, we can choose equation (4)′ as the free energy of the ferroelectric materials.

In the present study, we adopt the equation (4)′ as the free energy to simulate the spontaneous polarization in ferroelectric materials under an applied electrical and mechanical field. The TDLGD equation has been adopted as the temporal evolution equation which governing the evolution of the domain structure [17]:

$$\frac{\partial P_i(\bm{x},t)}{\partial t} = -L\frac{\delta F}{\delta P_i(\bm{x},t)} \quad (i=1,2,3) \tag{7}$$

where $P_i$ is used as the order parameters, $L$ denotes the kinetic coefficient. $F$ is the total free energy of the system, $\delta F/\delta P_i(\bm{x},t)$ represents the thermodynamic driving force for the spatial and temporal evolution of the simulated system. The applied electric field is $E_0\sin(2\pi ft)$. Substitutes equation (4)′ into equation (7), the kinetic equation of the evolution of the spontaneous polarization driven under the applied electric field can be expressed as

$$\frac{\partial \bar{P}}{\partial t} = -L[\Re \cdot \tanh(\Re \cdot \bar{P}) - E_0\sin(2\pi ft)] \tag{8}$$

According to equation (8), four typical kinds of hysteresis loops in ferroelectric materials can be numerically obtained by employing a Runge-Kutta method and by adjusting the synthetically parameter ($\Re$) employed in the model(Fig.2). It should be also noted that the differential equation (8) have two solutions in the real number space since it is a super equation/function containing the tanh(x) function. These four type loops have also been summarized in reference [20]. The slim loop(Fig.2a) and classical loop(Fig.2d) are the most popular loops and have been reported extensively. The frequency dependence(Fig.2d) predicted by the model is consistent with the experiment results[25]. The double loop (Fig.2c) and triple loop(Fig.2b) have been observed experimentally in antiferroelectric ceramics in 1953 [26] and in 1956 [27].

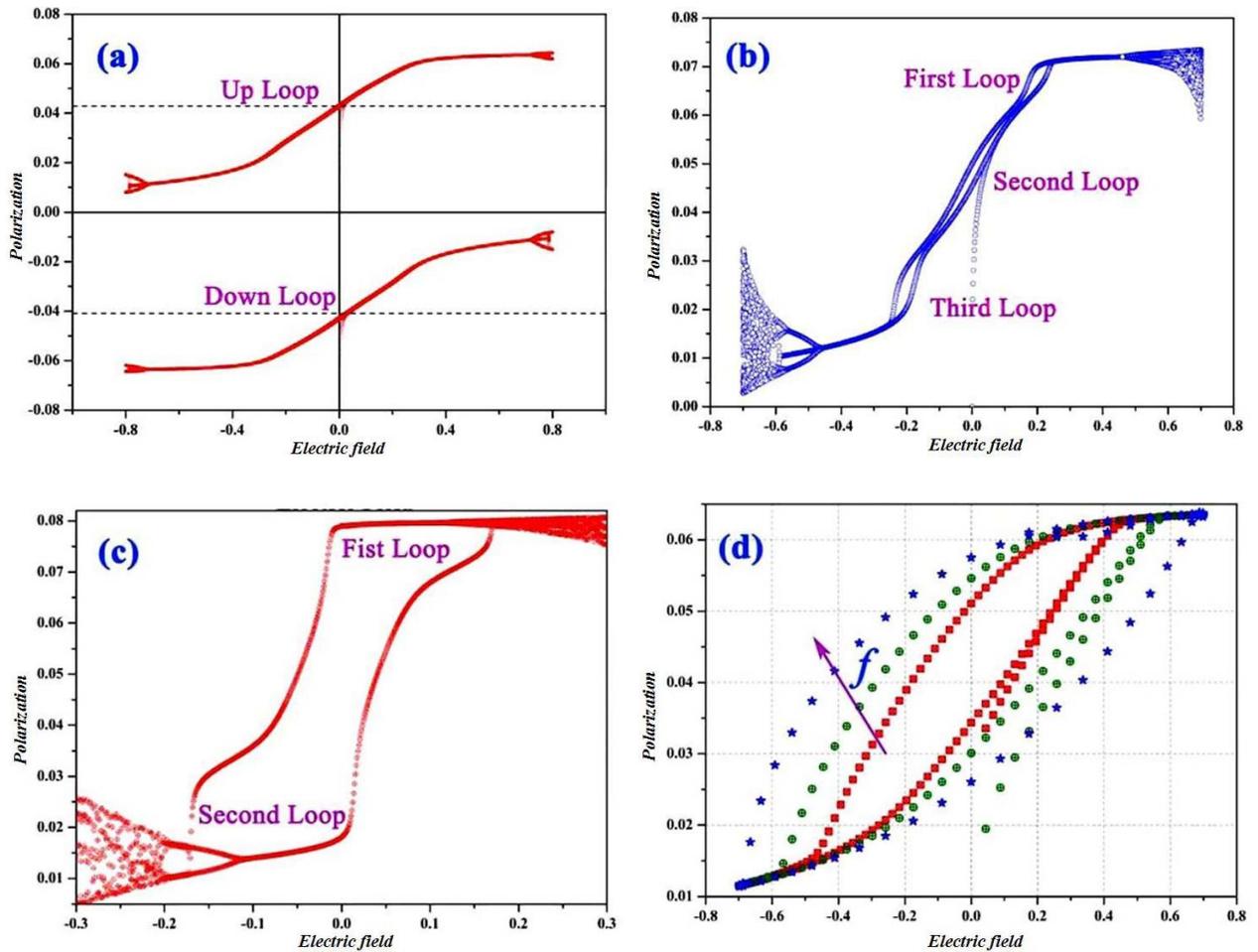

Fig. 2 Four patterns of hysteresis loops depicted by the model
(a) slim loop (b) triple loop (c) double loop (d) classical loop

## 4. Summary

Majority of the phenomenological models related to the hysteresis loops of ferroelectric materials is purely dependent on electric field. In order to understand the physical hysteresis loops clearly, we constructed a new model which is combined with the electric, the temperature, and the stress as a synthetically parameter in this paper. By comparing the predicted results with the experimental data in literatures, it renders that the simple and promising model derived in this paper can describe the experimentally diversified patterns of hysteresis loops in ferroelectric materials through adjusting the external excitation, or the synthetically parameter (e.g., electric, temperature, and stress, etc.) employed in the model.

**Acknowledgements**

This work was financially supported by the Research Starting Funds for Imported Talents of Ningxia University (Grant No. BQD2012011), and partially supported by the Natural Science Funds of Ningxia (Grant No. NZ15034) and the National Natural Science Foundation of China (NSFC) under Grant number of 11662014.